\documentclass[12pt,preprint]{aastex}

\newcommand{\HI}{\mbox{H\,{\sc i}}} 
\shorttitle{Tully-Fisher for PRGs}
\shortauthors{Iodice et al.}

\begin{document}

\title{Polar Ring Galaxies and the Tully Fisher relation: implications 
for the dark halo shape}

\author{E. Iodice, M. Arnaboldi}
\affil{INAF - Osservatorio Astronomico di Capodimonte, via Moiariello 16, 80131
Napoli, Italy}
\email{iodice@na.astro.it, magda@na.astro.it}
 
\author{F. Bournaud, F. Combes}
\affil{Observatoire de Paris, LERMA, 61 Av. de l'Observatoire, F-75014, Paris, 
France}
\email{frederic.bournaud@ens.fr, Francoise.Combes@obspm.fr}

\author{L. S. Sparke}
\affil{University of Wisconsin, Department of Astronomy,
475 N. Charter St., Madison, WI 53706-1582, U.S.A.}
\email{sparke@astro.wisc.edu}

\author{W. van Driel}
\affil{Observatoire de Paris, Section de Meudon, GEPI, CNRS FRE 2459 , 
5 place Jules Janssen, F-92195 Meudon Cedex, France}
\email{Wim.vanDriel@obspm.fr}

\author{M. Capaccioli}
\affil{INAF - Osservatorio Astronomico di Capodimonte, via Moiariello 16, 80131
Napoli, Italy}
\email{capaccioli@na.astro.it}


\begin{abstract}
We have investigated the Tully-Fisher relation for Polar Ring 
Galaxies (PRGs), based on near infrared, optical and \HI\ data available for a 
sample of these peculiar objects. 
The total K-band luminosity, which mainly comes from the central host galaxy, 
and the measured \HI\ linewidth at $20\%$ of the peak line flux density, 
which traces the potential in the polar plane, place most polar rings 
of the sample far from the Tully-Fisher relation defined for spiral galaxies, 
with many PRGs showing larger \HI\ linewidths than expected for the observed 
K band luminosity. 
This result is confirmed by a larger sample of objects, 
based on B-band data. 
This observational evidence may be related to the dark halo shape and 
orientation in these systems, which we study by numerical 
modeling of PRG formation and dynamics: the larger rotation velocities 
observed in PRGs can be explained by a flattened polar halo, aligned with 
the polar ring.
\end{abstract}

\section{Introduction}\label{intro}

Polar Ring Galaxies (PRGs) are peculiar objects composed of a central 
spheroidal component, the host galaxy, surrounded by an outer ring, 
made up of gas, stars and dust, which orbits nearly perpendicular to the plane 
of the gas-poor central galaxy (Whitmore et al. 1990).
Previous papers (Arnaboldi et al. 1995; Arnaboldi et al. 1997, Iodice et al. 
2002a, 2002b, 2002c) found that even where the morphology of the host galaxy 
resembles that of an early-type system, PRGs show many similarities with 
late-type galaxies.
The PRGs are characterised by a large amount of neutral hydrogen (\HI), always
associated with the polar structure (Schechter et al. 1984; van Gorkom et 
al. 1987;  Arnaboldi et al. 1997), and by a gas-to-total luminosity ratio
in the B-band typical of late-type galaxies.

The connection between PRGs and spirals is important for
two of the best studied systems: NGC 660 (van Driel et al. 1995) 
and NGC~4650A, for which the new surface photometry
(Iodice et al. 2002a; Gallagher et al. 2002), based on near-infrared (NIR) 
and optical Hubble Space Telescope (HST) data, have confirmed that the polar 
structure appears to be a disk of a very young age. 
By exploring the properties of the host galaxy and ring in the optical 
and NIR, for a sample of PRGs, Iodice et al. 
(2002a, 2002b, 2002c) found that the connection with spirals
is tighter. Iodice and collaborators found that in almost all PRGs {\it i)}  
the host galaxy is bluer  (even NIR and optical colors are similar to those of 
late-type galaxies) and younger (age vary between 1 and 5 Gyrs) 
than normal early-type galaxies, 
and it is characterised by a ``compact'',  nearly-exponential bulge, 
with a brighter and smaller disk than those found in standard S0 galaxies;
{\it ii)} the polar structure is even bluer than the host galaxy, 
and it has colors similar to those of dwarf irregular and spiral galaxies. 

The Tully-Fisher relation (TF) is the most important scaling relation 
for disks (Tully \& Fisher, 1977): this is an empirical relationship 
between the disk rotational velocity ($V_{rot}$) and its absolute 
luminosity ($L$), where $L\propto V_{rot}^4$, approximately. 
The TF is extensively used to estimate extragalactic 
distances (e.g. Sakai et al. 2000; Tully \& Pierce 2000, and references 
therein). Furthermore, this relation is considered a critical constraint 
in galaxy formation theories (Dalcanton et al. 1997; McGaugh \& de Blok 1998; 
Mo et al. 1998; Steinmetz \& Navarro 1999, van den Bosch 2000).
Several studies have addressed the origin of the TF relation
(Kauffmann, White \& Guiderdoni 1993; Cole et al. 1994; Silk 1997; 
Avila-Reese, Firmani \& Hernandez 1998, Heavens \& Jiminez 1999;
Elizondo et al. 1999; van den Bosch, 2000), but a definitive conclusion 
is yet to be found. 

In the past few years, several studies have asserted the validity of 
the TF relation for some classes of disk galaxies which show different 
photometric and kinematical properties with respect to `classical', 
high-surface-brightness spiral galaxies 
(Matthews, van Driel \& Gallagher 1998a, 1998b; McGaugh et al. 2000; 
Chung et al. 2002). In particular, the studies by Matthews et al. (1998b) 
and McGaugh et al. (2000), while exploring the TF relation for 
extreme late-type spiral galaxies and for faint field galaxies,
indicated  the validity of 
the ``baryonic TF relation'', which accounts for the gas mass 
in addition to the mass of the optical component by making a 
{\it baryonic correction} (Milgrom \& Braun 1988).
These latest developments indicate that the TF relation is 
probing a very close liaison between the dark halo parameters and the
total quantity of baryons in galaxies: the dark halo, which is
responsible for the \HI\ linewidth and the flat rotation curve in the outer
regions of a disk, is tuned to the total amount of baryons in the 
luminous component.

In PRGs, the \HI\ linewidth ($\Delta V$) measures the dynamics along the 
meridian plane, which is dominated by the dark matter, while 
the baryons are nearly equally distributed between the host galaxy and the 
polar ring. 
We wish to investigate the position of the PRGs in the $\log(\Delta V)-L$ 
plane, and study via N-body simulations of 3-D systems whether the
dark halo shape may influence their position in the $\log(\Delta V)-L$ plane,
with respect to the TF relation of bright disks.
The question of the dark halo shape is important {\it i}) to constrain 
dark matter models, through cosmological simulations 
(Navarro, Frenk \& White 1996, 1997; Bullock et al. 2001) which predict the 
distribution of the halo shapes and 
the universal radial dependence of the dark matter distribution; 
{\it ii}) to give hints on the nature of dark matter (see Combes 2002 
as a review); and furthermore {\it iii)}
the dark halo properties in PRGs can give important constraints 
on the formation scenarios for these peculiar objects, which is still 
an open issue (see Iodice et al. 2002a and 2002c and references therein).

In Section~\ref{obs} we present the NIR, optical and HI observations 
available for a sample of PRGs and discuss the PRGs 
location in the $\log(\Delta V)- L$ plane; and we derive some insights on the 
dark halo shape in these objects via N-body, 3- and 2-D models 
in Section~\ref{DM}. 
In Section~\ref{discu} we evaluate the uncertainties and discuss our
findings with respect to relevant dark halo parameters, i.e. total dark mass,
central density and flattening , and then derive our conclusions in 
Section~\ref{concl}.

\section{Observations}\label{obs}
\subsection{Near-infrared, B band, and \HI\  observations of polar ring galaxies}

New near-infrared J, H and Kn images are available for a sample of PRGs 
(Iodice et al. 2002a, 2002b, 2002c), which are selected from the Polar Ring 
Catalogue (Whitmore et al. 1990), and for ESO~235-G58, which was classified as 
a PRG related object by Buta \& Crocker (1993). 
Data were obtained at the 2.3 m telescope of the Mt. Stromlo 
and Siding Spring Observatory, with the CASPIR infrared camera 
(McGregor 1994).
The angular resolution of this camera is 0.5 arcsec pixel$^{-1}$ and it has
a field of view of $2.0' \times 2.0'$.
A detailed description of the data reduction and analysis is given by 
Iodice et al. (2002a, 2002b, 2002c).
The average photometric error on K magnitudes is 0.05 mag, and a detailed
discussion is given in Iodice et al. (2002b).
This sample contains only 6 PRGs, while more data are available in the 
optical bands. Since we wish to investigate the validity of the TF relation 
for PRGs, it is worthwhile to compare the PRG properties with those of a very 
large sample of spiral galaxies. 
To derive the TF relation for normal disk galaxies, we will use the very 
large and detailed dataset available in the I-band from Giovanelli et al. 
(1997). 
The B band magnitude is known for many PRGs (Whitmore et al. 1990; 
Van Driel et al. 2000a, 2002b; Gallagher et al. 2002), and 
we estimate the B band magnitude for spiral disks in the sample from 
Giovanelli et al. using the observational relation between morphological 
type index and the $B-I$ colors (de Jong 1996). 

In PRGs the bulk of the light comes from the host galaxy, which is a nearly 
dust free component in most systems (see Iodice et al. 2002a, 2002b).  
In most PRGs of our sample, the central most luminous part of the host galaxy 
is not obscured by the dust in the PR region which passes in front of it, so 
there is no significant correction for internal extinction. For the 3 PRGs 
(ESO 603-G21, ARP 230, ESO 235-G58) where the dust associated with the PR 
obscures the central regions of the host galaxy, we have corrected the K and 
B total magnitudes for internal extinction, following the equations from 
Tully et al. (1998). We expect little extinction within the polar ring itself,
since this component resembles low-surface-brightness disk galaxies 
(see the HST images in Iodice et al. 2002a and Gallagher et al. 2002), where 
dust extinction is not high (Matthews and Wood 2001).

The \HI\ integrated line profile data were obtained from several published 
hydrogen observations of PRGs, carried out by e.g. Richter et al. (1994), 
van Gorkom et al. (1987), van Driel and collaborators (2000, 2002a, 2002b), 
with several radio telescopes.

The absolute Kn magnitude ($M_{Kn}$), B band absolute magnitude, and measured 
linewidth at $20\%$ of the peak line flux density ($\Delta V_{20}$), for 
each PRG in our sample are listed in Table~\ref{tab1}. 
For all distance-dependent quantities we have assumed 
$H_0=75$ km s$^{-1}$ Mpc$^{-1}$.

\subsection{PRGs and the TF relation for spiral galaxies}
We wish to compare the position of the PRGs in the $\log(\Delta V)-L$ 
plane with the TF relation observed for bright spiral galaxies. 
In principle, we do not know what expect, 
because the two quantities, rotational velocity and absolute 
luminosity, are related to two different components in the PRG, the polar 
structure and the host galaxy respectively, which lie in two decoupled 
planes. In fact the total K light comes mostly from the central galaxy 
(Iodice et al. 2002a, 2002b, 2002c),
whereas the observed \HI\ linewidth traces the dynamics
of the polar structure along the host galaxy's meridian plane.

In Figure~\ref{fig1} we show the K-band TF relation 
for a sample of spiral galaxies studied by Verheijen (1997, 2001). 
The values of $M_{K}$ and $\log(\Delta V_{20})$ for PRGs are also shown on 
this plot.  Our and Verheijen's data sets have very similar
photometric properties and limiting magnitudes.\\
We see that five PRGs lie near to the high-velocity boundary
of the TF relation, 
or show larger velocities (for a given luminosity) than disk galaxies. Only 
the PRG AM2020-504 shows a lower rotation velocity for its K-band 
absolute magnitude.\\

In the B-band, the tendency of PRGs to have larger velocities with respect 
to the TF is confirmed for a larger sample of PRGs compared with data for 787 
disk galaxies (Giovanelli et al. 1997). In Figure~\ref{fig2}, we see that two 
PRGs lie at lower velocities than those predicted by the TF for spiral 
galaxies, two objects lie on the TF relation, and twelve PRGs either lie 
on the high velocity boundary of the TF relation or show much larger 
velocities. Van Driel et al. (2002b) also find that most kinematically 
confirmed PRGs show larger \HI\ profile widths than bright spiral galaxies, 
at a given luminosity.

\section{The PRG positions in the $\log(\Delta V) -L$ plane 
and their implications for the dark halo}\label{DM}

\begin{figure}
\includegraphics[angle=270,width=16.3cm]{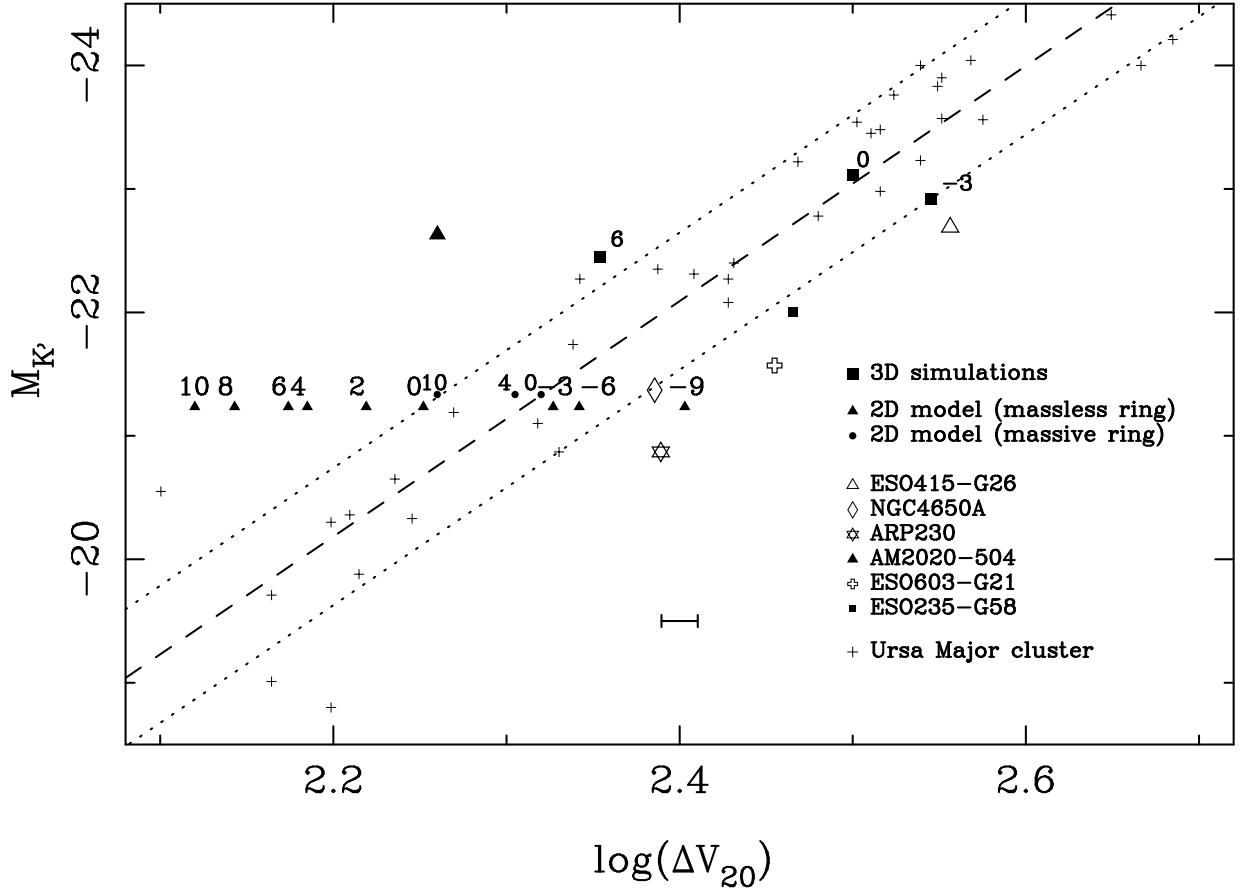}
\caption{Absolute magnitude in the K band vs. the measured \HI\ linewidth 
at $20\%$ of the peak line flux density ($\Delta V_{20}$), 
for PRGs of the selected sample, compared with a sample of spiral galaxies 
from Verheijen (2001), and with the results from 3-D simulations and 2-D 
models (for massless rings and rings that are as massive as the host galaxy).
We have assumed $H_0 = 75$ km s$^{-1}$ Mpc$^{-1}$, as made also by Verheijen.
The long-dashed line is a linear interpolation of the TF relation for spiral 
galaxies, and the short-dashed lines show the width at 15\% of the peak of 
the statistical distribution of spiral galaxies. For both 
models, the flattening of the halo is indicated  next to each circle (massless 
ring) or triangle (very massive ring): a positive $x$ number indicates that 
it is an E$x$ 
halo with an equatorial flattening, while $-x$ corresponds to an E$x$ halo 
flattened toward the polar plane. The results from the 3-D models 
shown in this plot are those computed for the accretion scenario 
(Reshetnikov \& Sotnikova, 1997) ; 
our values for $\Delta V_{20}$ vs. M$_K$ are very similar when one considers 
the merging scenario (Bekki, 1998). \label{fig1}}
\end{figure}

\begin{figure}
\includegraphics[angle=270,width=16.3cm]{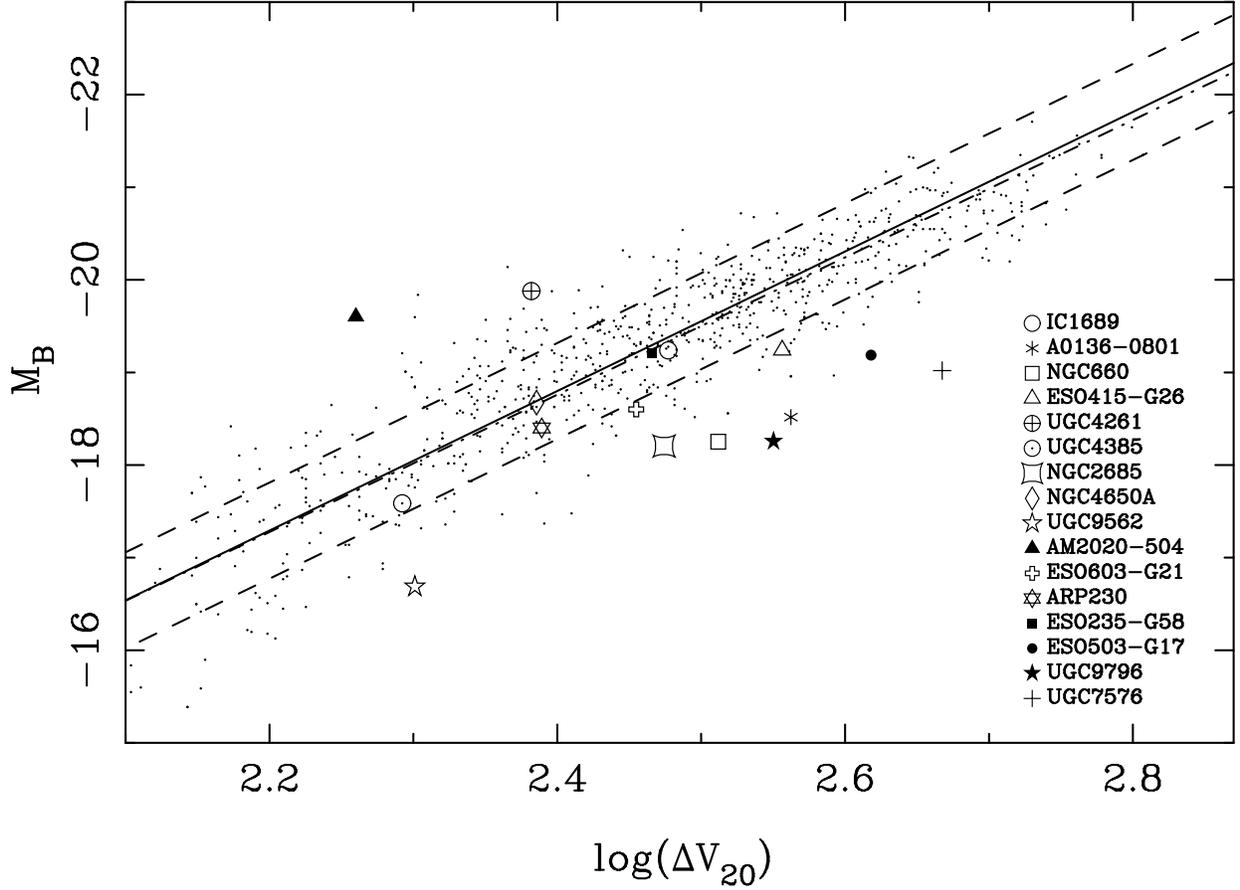}
\caption{\label{fig2} Absolute magnitude in the B band vs. the linewidth 
at $20\%$ of the peak line flux density ($\Delta V_{20}$), for 15 PRGs. 
Data for disk galaxies (dots) are from Giovanelli et al. (1997). 
Absolute magnitudes have been normalised to the same value of $H_0$ for 
PRGs and disk galaxies ($75$ km s$^{-1}$ Mpc$^{-1}$). 
A linear interpolation of the TF relation is shown for these disk galaxies 
(solid line), $81\%$ of which lie inside the dashed lines that are 
computed at 25\% of the peak of the statistical distribution of spiral 
galaxies. The long-short dashed line is  obtained for unbarred disks, 
seen nearly edge-on ($i\geq80°$, see Section~\ref{discu}).}
\end{figure}

\begin{figure}
\includegraphics[angle=0,width=5.5cm]{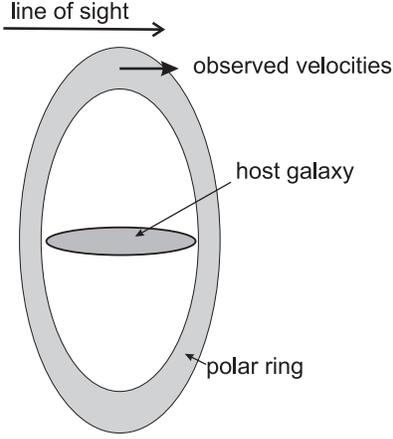}
\caption{\label{schem}Line of sight for most PRGs: both components are seen 
nearly edge-on, thus the measured velocities in the polar structure are the 
smallest ones when the ring is eccentric, i.e. when the potential is 
flattened along the equatorial plane. This line of sight corresponds to the 
PRGs of our sample, and it is adopted for the analysis of the numerical 
models.}
\end{figure}

\begin{figure}
\includegraphics[angle=270,width=6.5cm]{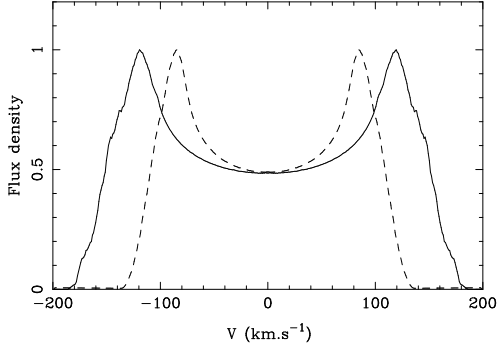}
\caption{\label{ecc} Simulated \HI\ profiles for circular and eccentric polar 
rings seen edge-on, as shown in Figure~\ref{schem}. 
Solid line: \HI\ profile computed for a circular ring, extending from 10 to 
15 kpc. Dashed line: \HI\ profile computed for an eccentric ring, 
with the same radial extension and radius, and ellipticity 0.35 (similar to 
the ring from 3-D simulations shown in Figure~\ref{mod3D}).
For gas, we assume a velocity dispersion of 10 $km/s$.
The values of $\log(\Delta V_{20})$ are 2.51 for the circular ring and 
2.38 for the eccentric ring. 
This difference is in agreement with the set of values for 
$\Delta V_{20}$, at a given total K or I band luminosity, 
obtained from PRG numerical models (see Figure~\ref{fig1}).}
\end{figure}

Observing higher or lower velocities with respect to the linear TF of 
disk galaxies is relevant for the discussion on the dark halo shape.
Via analytical models and simple assumptions about the mass distribution, 
either luminous or dark, we can estimate where the PRGs ought to lie
in the $\log(\Delta V)-M_{K/B}$  plane (Fig.\ref{fig1} and Fig.\ref{fig2}),
with respect to the TF relation for disk galaxies.
If there were no dark matter, and the gravitational potential 
were oblate in the same sense as the flattened host galaxy, the polar ring 
would acquire an eccentric shape. When the polar ring 
and the host galaxy are both seen edge-on, which is close to being the case
for most of our PRGs, the net effect will be that the line-of-sight (LOS)
polar ring velocities are reduced, see Fig.~\ref{schem}. 
In the logarithmic, scale-free, 
potential case, a simple formula gives the expected velocity ratio between 
the major and the minor axis components as:
\begin{equation}
\epsilon_v = 1 - \frac{v_{major}}{v_{minor}} = 
\epsilon_\rho \simeq 2\epsilon_\Phi
\end{equation}
from Gerhard \& Vietri (1986), where $\epsilon_\rho$ is the flattening 
(1-axis ratio) of the density distribution and $\epsilon_\phi$ is the 
potential flattening. 
Therefore we would expect PRGs to have on average lower velocities with 
respect to what would be measured in the equatorial plane: see Fig.\ref{ecc}.
This implies that when the polar structure is eccentric, the observed LOS 
velocities in $\Delta V_{20}$ are the smallest, i.e. those from the 
particles in the polar regions, as shown in Figure~\ref{ecc}. 
Thus the observed $\Delta V_{20}$ depends on both the mean velocity
along the ring, and the ring eccentricity.
On the contrary, Fig.\ref{fig1} and Fig.\ref{fig2} show 
that the majority of PRGs 
have larger velocities than expected in the $\log(\Delta V)-M_{K/B}$
plane. Therefore we need to investigate how these velocities can 
be produced, and how they may depend on the intrinsic properties of
the dark galaxy halo.

\begin{figure}
\includegraphics[angle=270,width=7cm]{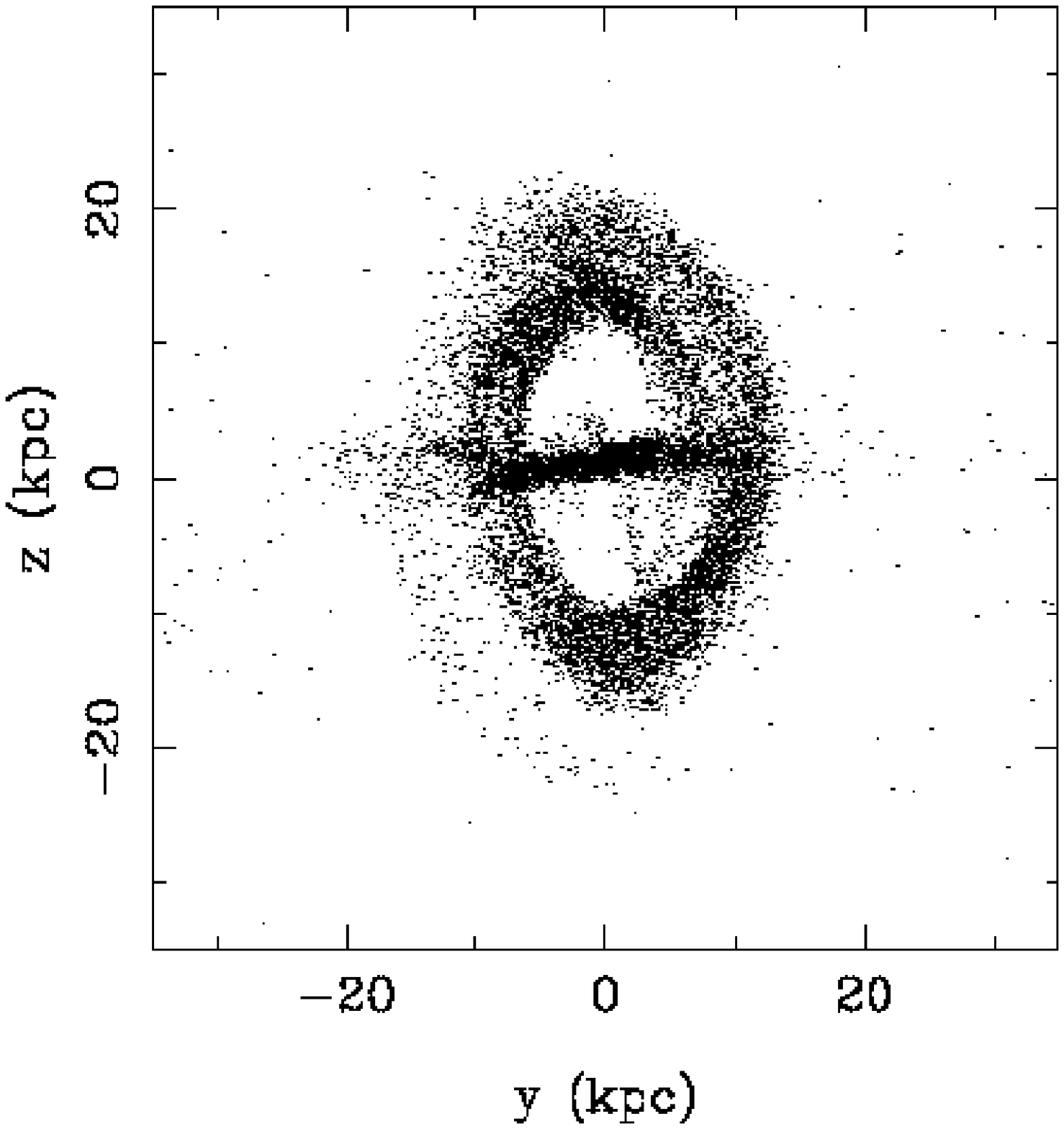}
\hspace{.3cm}
\includegraphics[angle=270,width=7cm]{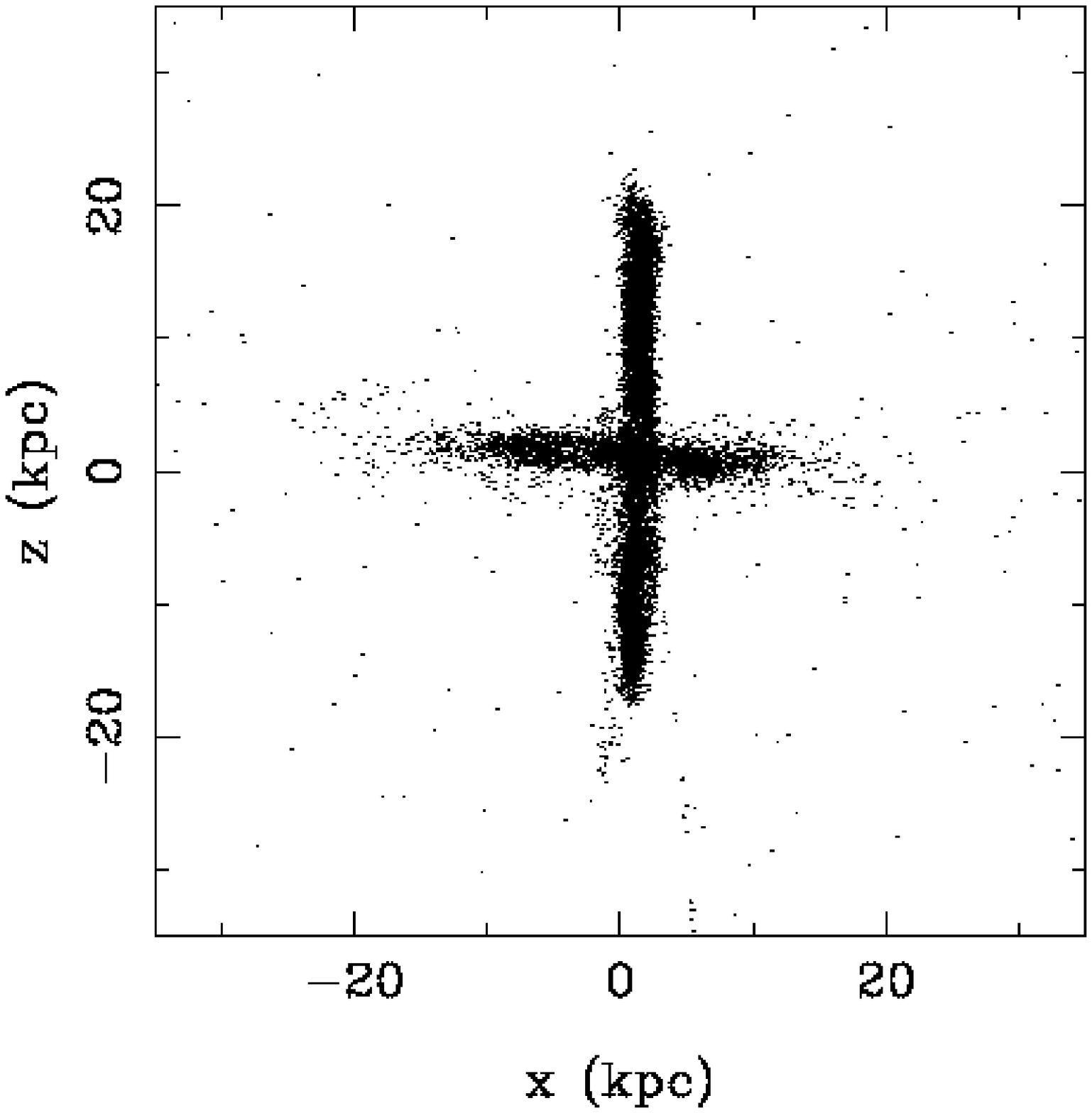}
\caption{\label{mod3D}Example of polar ring formed in the 3-D N-body model, 
according to an accretion scenario. In most observed PRGs both the central 
disk and the polar ring are seen nearly 
edge-on (right panel). This polar ring is eccentric, since the gravitational 
potential is flattened by the host disk and an E4 dark halo.}
\end{figure}

We first compute a series of N-body models of the formation of polar rings 
in galaxy mergers or by tidal accretion of gas from another galaxy
(Bournaud \& Combes 2002). Both scenarios are known to form polar structures 
(see Reshetnikov \& Sotnikova 1997; Bekki 1997, 1998). 
The numerical code computes the gravitational potential via an FFT method 
(James 1977), and includes sticky particles gas dynamics (Schwarz 1981), 
and star formation that is assumed to obey a generalised Schmidt law 
(Schmidt 1959). Both scenarios for formation of polar rings give similar 
results for the kinematics of the ring after its formation. The details of 
the N-body models and numerical codes are described in
Bournaud \& Combes (2002).
Fig.\ref{mod3D} shows an example of the structures produced.

These 3D N-body models do not allow us to choose the dark halo shape after 
the polar ring is formed. Moreover, we cannot separate the influence of the 
host galaxy potential, the dark halo flattening, and the ring self-gravity, 
which are all important in order to understand what our observations may 
imply for the PRG structure. 
We therefore developed a simpler model, where every parameter can be varied. 
In this model, the dark halo and the host galaxy are two rigid potentials. 
The eccentricity of the polar structure is given by the shape of the 
particle orbits in this rigid potential. 
Particles are launched around the polar plane and are followed for 4 Gyrs, 
before their kinematics is analysed. 
The polar ring self-gravity may also be included via a 2-D FFT algorithm that 
computes the gravitational potential in the polar plane.

Our N-body 3D models of polar rings (see Figure~\ref{mod3D}) show 
different characteristics depending on the shape of their dark halo.
We see that when the halo is oblate and flattened towards the host galaxy, 
the observed velocities are then smaller, and the PRGs lie on the left side 
of the TF relation of high-surface-brightness disks.
When the oblate halo is flattened towards the polar ring, 
the observed velocities are larger, shifting the PRGs to the right side of the 
diagram. As most of the observed PRGs have larger velocities 
compared to the TF relation for spiral galaxies, it seems to suggest that 
their dark halo is flattened towards the polar plane. 

The second model is initially computed for massless rings. 
We choose a 10kpc-scaled host galaxy, and a polar ring whose mean radius is 
15 kpc. The dark halo is computed to produce a flat rotation curve at large 
radii, its mass is about 0.8 times the host mass inside 10 kpc radius, 
and 1.4 times the host galaxy mass inside 15 kpc. 
A halo flattened towards the equatorial plane reduces the measured spread in radial
velocity in the polar structure: the more flattened the potential, the more eccentric 
the ring is and the smaller the observed velocities are.
When the halo is spherical, 
the potential is flattened by the host galaxy luminous mass, therefore
the ring is still eccentric, and the PRG velocities then lie on the
low-velocity boundary of the TF relation, far from the observational
result. This is what was found by Combes and Arnaboldi (1996) in their 
detailed model for the polar ring galaxy NGC4650A: a spherical
halo cannot reproduce the large ring velocities, because the ring is 
eccentric, unless when the halo is very massive (but this case is
easily ruled out, see Sect.~\ref{MDM}).

\begin{figure}
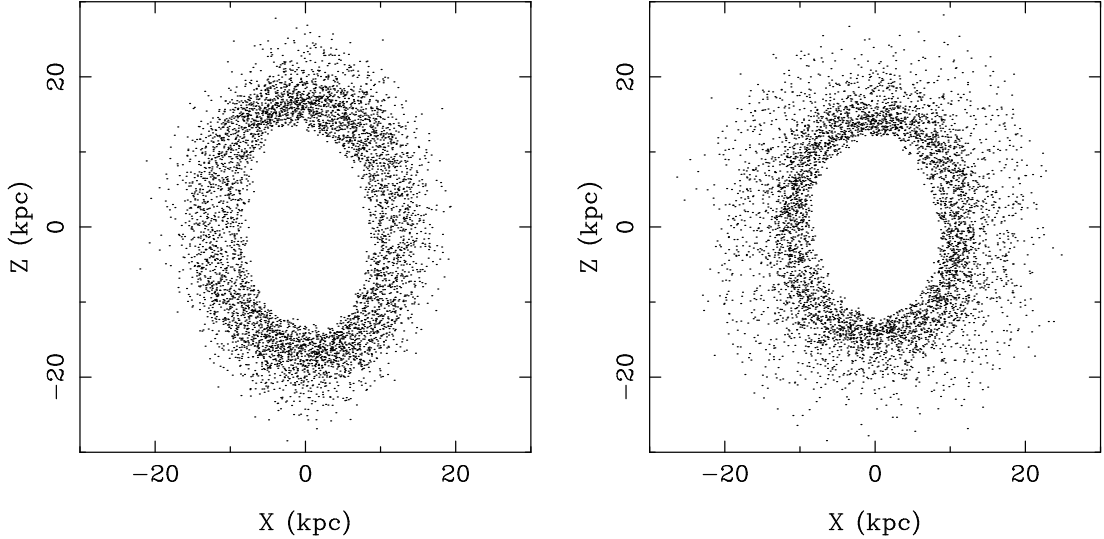

\includegraphics[angle=270,width=7cm]{iodicee_fig6A.ps}
\hspace{.3cm}
\includegraphics[angle=270,width=7cm]{iodicee_fig6B.ps}
\caption{\label{mod2D}Polar rings from our 2-D model, formed during an 
accretion event, obtained with a E3 halo flattened in the equatorial plane
($x-y$). The ring lies in the $z-x$ plane.
The massless ring (on the left) is eccentric, which reduces the observed 
velocities.
Though the very massive ring (on the right, the ring is as massive as the host 
galaxy) is rounder, its self-gravity is not sufficient to reproduce 
the observed velocities, unless the halo is also flattened towards the ring.
The measured linewidth  $\Delta V_{20}$ is $30\%$ larger for the massive 
polar ring than for the massless ring.}
\end{figure}

We now wish to account for the ring self-gravity: the ring mass for 
all objects in our sample is believed to be a significant fraction of the 
stellar mass of the host galaxy or of the same order, as it is in the case for 
NGC~4650A (Iodice et al. 2002a). 
When the halo is spherical or flattened towards the host galaxy equatorial plane, 
its self-gravity makes the polar ring more circular (see Figure~\ref{mod2D}).
In Figure~\ref{fig1} we have shown the effects 
for polar rings whose mass is equal to the mass of the host galaxy. 
When the polar ring is lighter, these effects are 
smaller; for instance when the polar mass is half the  
mass of the luminous HG, we find results that are comparable to those obtained 
with the first model.
When self-gravity is considered, the simulated PRGs lie near 
the TF relation for bright disk systems of Fig.\ref{fig1}. 
And yet, when the flattened dark halo is aligned with the host galaxy equatorial 
plane, the ring's self-gravity cannot account for the large velocities observed in 
most polar rings, even when the ring is as massive as the host galaxy. 
But, when dark halos are flattened towards the polar structures, the ring
self-gravity is less important, because the flattened potential
already makes rings more circular: the positions of the corresponding
points does not move much from the the massless ring case, when the
ring mass is increased, see Figure~\ref{fig1}. 
Indeed such systems seem able to explain the large velocities observed in 
the polar structures, and confirms the conclusions drawn from our 3-D 
accretion and merging models.

Let us now consider the case of AM~2020-504 and UGC~4261, which show 
lower velocities than bright disks, for their observed K and B band
absolute magnitudes. 
Both galaxies have narrow rings with very small masses, therefore they are 
well described by our massless ring model. This model easily predicts the 
observed velocities in the polar ring, when the potential is significantly 
flattened towards the host galaxy equatorial plane. 
This is confirmed in the case of 
AM~2020-504 by the photometric study of the host galaxy, which is a flattened 
elliptical (E4, see Arnaboldi et al. 1993). 
Moreover, if those systems are embedded in massive dark halos that contribute 
much to the potential near the polar ring, these halos must be flattened too.

\section{Discussion}\label{discu}

In the $\log(\Delta V) - M_{K/B}$ planes shown in Fig.~\ref{fig1} and 
Fig.~\ref{fig2}, most 
PRGs have larger \HI\ rotation velocities than standard spiral galaxies, 
at a given K or B-band luminosity of the stellar component.
Our N-body simulations have suggested that a likely explanation for this
effect is a flat dark halo, whose main plane is 
aligned with the host galaxy meridian plane, and prevents the polar
ring to become eccentric. 
The question arises if other effects, i.e. non-homogeneities in the TF 
relations for spirals, caused by bar and/or non edge on disks, or
larger M/L ratios, can produce similar results and therefore be 
alternative explanations for the high velocities observed in PRGs.

\subsection{Possible biases in the spiral sample}
PRGs are peculiar objects that are easily identified when both their 
central component and polar ring are observed nearly edge-on. 
When dealing with the TF relation for spiral galaxies, 
we have assumed that it was valid for axisymmetric edge-on disks. 
Yet, the large sample from Giovanelli et al. (1997) contains barred disks and 
disks at lower inclination with respect to the LOS. 
Bars may cause a bias in the observed velocities, and calculating
the real value of $\Delta V$ for a disk at a lower inclination 
cannot be done exactly.
We have therefore selected from the Giovanelli et al. sample those disk 
galaxies that are not barred and are seen nearly edge-on ($i\geq 80°$); 
we built a new TF relation from this new sample of 179 disks, which
is now free from those biases that may cause lower \HI\ velocities.
The linear interpolation of this new TF relation is shown 
in Figure~\ref{fig2} : it is still very close to the one of the original sample
and indicates that barred disks and disks with lower inclination 
do not modify the mean TF relation (although the scatter is smaller, for bars 
increase the scatter of the TF relation).
So this cannot explain why PRGs have larger observed \HI\ velocities
than bright disks.

\subsection{The baryonic Tully-Fisher relation}
\begin{figure}
\includegraphics[angle=270,width=8cm]{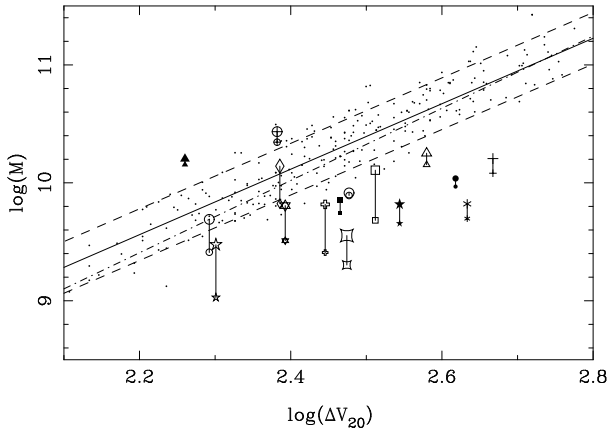}
\caption{\label{massTF}Plot of the stellar, mass in solar units, 
(smaller symbols for PRGs, long-short dashed line for spirals) and 
the stellar plus gaseous mass (solid line and dots for spirals, large symbols 
for PRGs) against the observed rotation velocities. Due to the similarities 
between host galaxies in PRGs and early-type spiral galaxies (Arnaboldi et 
al. 1995; Iodice et al. 2002a, 2002b), we have assumed
$M/L=2$ for this component, as for spirals (de Jong 1996).
In this plot, data for disk galaxies are from Mathewson \& Ford (1996).}
\end{figure}

We have already mentioned (Sec.~\ref{intro}) that PRGs are gas-rich, thus we 
wish  to investigate whether once we account for the gas mass, this might 
move them back  on the TF  relation, i.e. if they are close to the baryonic TF 
relation  (Matthews et al. 1998b;  McGaugh et al. 2000). 
A mass-to-velocity TF relation for PRGs is shown in Figure~\ref{massTF}:
PRGs are now closer to the TF relation for disk galaxies, 
but still do not lie on this relation: i.e. the higher velocities in the PR 
cannot be explain by additional mass in the \HI\ associated within the PR 
itself.

\subsection{Host galaxy mass-over-light ratios}
Another possible explanation for higher \HI\ rotation velocities is that PRGs 
have larger mass-to-light ratios than spiral disks of similar K\&B-band 
luminosity, i.e. for the same observed $\Delta V_{20}$, PRGs may be  
less luminous on average in K/B than standard spiral disks. 
Therefore we have computed the linear interpolation of the TF relation for 
the $Sa$ galaxies alone, of the Giovanelli et al. (1997) sample.
The average B-band magnitude of this sample corresponding to 
$\log(\Delta V)=2.5$ is -19.35, while it is brighter, -19.16,
for the whole Giovanelli sample. Still, the difference of 0.19 mags is not 
enough to place most PRGs on the TF relation, which would require
a value 5 times larger.

\subsubsection{Comparison between PRG host galaxies and S0s}
The similarities between PRGs and spiral galaxies, from NIR and
optical photometry, prompted us to compare PRGs with the TF for early-type 
spiral galaxies rather than with lenticulars. 
But we have also compared the position of PRGs in the $\log(\Delta V) - M_{B}$
plane with the TF relation for S0 galaxies\footnote{The TF relation
for S0 galaxies was obtained by comparing the asyntotic rotation
speeds of stars with the absolute magnitude (see Mathieu et
al. 2002).} 
(Neistein et al. 1999, Mathieu et al. 2002), and we find that many
PRGs lie near the TF relation for S0 galaxies. 
The offset between the TF relation for S0 galaxies and the 
TF for bright spiral galaxies is caused by the S0s being less luminous than 
bright spirals, at a given velocity. As suggested by Mathieu et al. (2002), 
this is consistent with the formation scenario in which S0 galaxies were 
formed from ordinary late-type spirals which were stripped of their 
star-forming medium.

\begin{figure}
\includegraphics[angle=270,width=8cm]{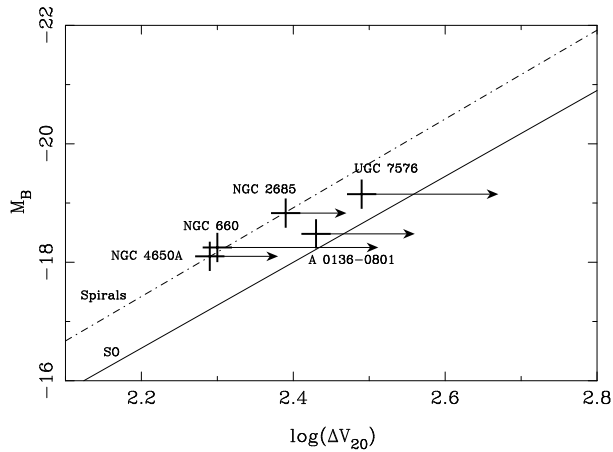}
\caption{\label{TFSO}TF relation for S0 galaxies 
(data from Neistein et al. 1999 and Mathieu et al. 2002) and PRGs.
The TF for spiral galaxies is plotted as a dashed line. 
Large crosses show the position of the central component in five PRGs. 
For NGC~4650A the circular velocity was derived from the rotation velocities 
and velocity dispersion of the stellar component (Combes \& Arnaboldi 1996), 
by computing the corresponding gas linewidth and assuming a 10 km/s dispersion 
(which is the standard value for disks). The total B magnitude for NGC~4650A
is from Iodice et al. 2002a. 
For NGC~660, velocities are derived from $H\alpha$ data (see van Driel et 
al. 1995). For NGC2586 and A0136-0801 the B magnitudes relative to the HG are 
given by Reshetnikov et al. 1994. The arrow shows the offset in 
$\log(\Delta V_{20})$ between the velocities in the host galaxy equatorial
plane and in the polar ring for each system. }
\end{figure}

Can this be a valid explanation for the PRGs position in the 
$\log(\Delta V) - M_{K/B}$ plane? If the PRG positions in this 
plane were caused by larger M/L ratios for the luminous component, 
as is the case for lenticular galaxies, 
it must be so for the host galaxy too, 
which is the component most similar to an S0 galaxy.
We have computed $\log(\Delta V)$  for
the host galaxies of the best studied systems, NGC~4650A, NGC~660, 
NGC~2685, UGC~7576 and A0136-0801, (Sackett et al. 1994; van Driel et al. 
1995; Simien \& Prugniel 1997; Whitmore et al. 1990; Schweizer et al. 1983), 
using optical absorption-line rotation curves along the host 
galaxy equatorial plane; these are shown in Fig.~\ref{TFSO}. 
The host galaxies in these five systems fall on the TF relation of
bright spiral disks, which indicates that {\it i)} the mass-to-light 
ratio of this component is different from those of standard S0s, 
as also found in recent studies of PRG NIR and optical colors 
(see Iodice et al. 2002a, 2002b, 2002c) and {\it ii)} a
luminous-to-dark matter content similar to those of standard bright disks. 

\subsection{Peculiar properties of the dark halo}\label{MDM}
Gerhard et al. (2001) showed that elliptical galaxies follow a TF relation
in the $\log(\Delta V) - M_*$, where $M_*$ is the total mass in the luminous 
component, which is shallower than the relation for spiral galaxies, 
even when the maximal $M/L_B$ is adopted to compute the total stellar masses.
This led Gerhard and collaborators (2001) to infer that elliptical galaxies have
slightly lower baryonic mass than spiral galaxies of the same circular
velocities, and that their dark halos are denser than halos of spiral galaxy 
with the same $L_B$. 
How much more massive must the dark halo be, to account for the velocities 
observed in polar rings?
In our numerical models, we had assumed a dark-to-visible mass ratio of about
1 inside the optical radius\footnote{For PRGs, the optical radius 
correspond to the ring average extension.}, 
as it is the case in spiral galaxies, and
only flattened polar halos could reproduce the observed \HI\ velocities.
The observed value of $\Delta V_{20}$ depends largely on the dark 
halo shape and the ring eccentricity, while it varies only as the square root 
of the total mass (and depends even less on the dark mass). 
Thus, a large amount of dark matter is needed when the halo 
is not polar: if one wishes to reproduce the PRG position on the TF
diagram with a spherical halo, then the dark mass needed is twice than 
the visible mass. With an E5 equatorial halo, the dark-to-visible mass 
ratio raises to about 3.5.\\
An increased total dark halo mass alone can explain the observed position 
of PRGs on the $\log(\Delta V) - M_{B}$  plane, but the required large 
dark masses are unrealistic: the fact that the PRGs 
host galaxies lie within the TF for spiral galaxies (see Fig.~\ref{TFSO})
indicates that the dark matter content in PRGs
is similar to that of bright spiral galaxies. 

Also in the spherical halo case, the host galaxy would be offset too from the 
TF relation of bright disks, because the dark mass. In the case of NGC~4650A, 
Combes \& Arnaboldi (1996) found that no 
dark matter was required to reproduce the host kinematics, and that the 
dark-to-visible mass ratio could not be as large as required by a spherical 
halo. If we want to account for observed velocities in PRGs with a spherical 
halo, we have to believe that the halo density drops more slowly than $1/r$.

A prolate dark halo associated with the host galaxy, i.e. with the major axis 
aligned with the HG major axis, 
could also account for the large velocities observed in rings, provided it 
is massive enough. To reproduce the large velocities in the PR, such a halo 
must be even more massive than for the oblate case. For an E5 prolate halo 
and a ring mass that amounts to 25 \% of the visible mass, we find that the 
dark-to-visible mass ratio inside the optical radius would be close to 4. 
Such a massive halo would cause a large offset of the host galaxy from the TF 
relation of bright disks, which is not observed. On the other hand, a prolate 
halo with a polar major axis can account for the velocities in PRGs and this
is one of the possibility of a halo flattened towards the polar ring.

Given the constraint from the host galaxy observed kinematics, the dark
halo shape alone can explain the high rotation velocities observed in 
polar structures, provided that the short axis of the halo lies 
perpendicular to the polar ring, in the equatorial plane of the central galaxy.
In the case of NGC~4650A, Combes \& Arnaboldi (1996) had already argued that 
a dark halo flattened towards the polar ring required less dark mass
to account for the large velocities observed in the polar ring, 
while the visible mass alone was enough to account for the velocities 
in the host galaxy.

\section{Conclusion}\label{concl}
In this paper we have explored the position of PRGs in the
$\log(\Delta V) - M_{K/B}$ plane with respect to the TF relation 
for spiral galaxies. The majority of PRGs show larger rotation velocities 
at a given luminosity than spiral galaxies. 
A comparison with the N-body simulations has led us to conclude that these 
large velocities indicate that the dark halo is most likely flattened towards 
the polar ring plane. This explains the position of 
PRGs with respect to the TF relation for normal spiral disks, 
the position of the host galaxy within the TF of bright disks,
and it may become an important constraint on the nature of dark matter.

\acknowledgments
The 3-D simulations in this work were computed on the Fujitsu 
NEC-SX5 of the CNRS computing center, at IDRIS. This research made use of 
the NASA/IPAC Extragalactic Database (NED) which is operated by the Jet 
Propulsion Laboratory, California Institute of Technology, under contract 
with the National Aeronautics and Space Administration.
M.A. wishes to thank O. Gerhard for many useful discussions on 
the TF relation for elliptical galaxies and for his comments on an
earlier version of this paper.



\clearpage

\clearpage

\begin{deluxetable}{llccccc}
\tabletypesize{\scriptsize}
\tablecaption{\label{tab1} Absolute Kn and B band magnitudes, 
heliocentric velocity, and \HI\ linewidth at 
$20\%$ of the peak line flux density for a sample of polar ring galaxies.} 
\tablewidth{0pt}
\tablehead{
\colhead{Object name} & \colhead{PRC} & \colhead{$V$} & \colhead{M$_{Kn}$} &
\colhead{M$_B$} & \colhead{$\Delta V_{20}$} & \colhead{Ref.}  \\
\colhead{} & \colhead{} & \colhead{km.s$^{-1}$} & \colhead{mag} & 
\colhead{mag} & \colhead{km.s $^{-1}$} & \colhead{}}
\startdata
A0136-0801 & A-01 &  5528 &         & -19.23 & 365 &  vD      \\
ESO415-G26 & A-02 &  4604 & -22.69  & -19.24 & 360 &  vD      \\
NGC2685    & A-03 &   871 &         & -18.20 & 335 &  vD      \\
UGC7576    & A-04 &  7022 &         & -19.02 & 465 &  vD      \\
NGC4650A   & A-05 &  2910 & -21.37  & -18.67 & 235 &  vD      \\
UGC9796    & A-06 &  5420 &         & -18.25 & 355 &  vD      \\
ARP230     & B-01 &  1710 & -20.95  & -18.40 & 245 &  vD      \\
IC1689     & B-03 &  4557 &         & -20.40 & 300 &  vG      \\
ESO503-G17 & B-12 & 10481 &         & -19.19 & 415 &  vD      \\
UGC9562    & B-17 &  1242 &         & -16.68 & 200 &  vD      \\
AM2020-504 & B-19 &  4963 & -22.63  & -19.60 & 182 &  vD,A    \\
ESO603-G21 & B-21 &  3124 & -21.74  & -18.60 & 285 &  vD      \\
NGC660     & C-13 &   850 &         & -18.25 & 325 &  vD      \\
UGC4261    & C-24 &  6408 &         & -19.87 & 240 &  vD      \\
UGC4385    & C-27 &  1969 &         & -17.58 & 195 &  vD      \\ 
ESO235-G58 &      &  4262 & -22.13  & -19.21 & 292 &  vD      \\
 \enddata
\tablenotetext{a}{References for values of $\Delta V_{20}$ : telescope codes 
(V:VLA, N: Nan\c{c}ay, G: Green Bank, P: Parkes); vG: van Gorkom et al. (1987),
 vD: van Driel et al. (2000, 2002a, 2002b and references therein) and 
Richter et al. (1994), A: Arnaboldi et al. (1993).}
\tablenotetext{b}{The B band magnitude for NGC~4650A is from Gallagher et al. 
(2002).}
\end{deluxetable}

\end{document}